\documentclass{article}

\usepackage{arxiv}

\usepackage[utf8]{inputenc} 
\usepackage[T1]{fontenc}    
\usepackage{hyperref}       
\usepackage{url}            
\usepackage{booktabs}       
\usepackage{amsfonts}       
\usepackage{nicefrac}       
\usepackage{microtype}      
\usepackage{lipsum}		
\usepackage{graphicx}
\usepackage{natbib}
\usepackage{doi}

\usepackage{adjustbox}
\usepackage{makecell}
\usepackage{verbatim}
\usepackage{tabularx}
\usepackage{amsmath}
\usepackage{multirow}

\title{Cascade learning in multi-task encoder-decoder networks for concurrent bone segmentation and glenohumeral joint assessment in shoulder CT scans}

\author{ \href{https://orcid.org/0009-0001-9738-0438}{\includegraphics[scale=0.06]{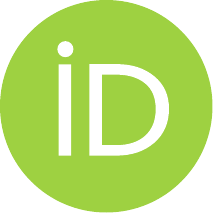}\hspace{1mm}Luca Marsilio} \\
	Department of Electronics,\\
	Information, and Bioengineering,\\
	Politecnico di Milano, Milan, Italy. \\
	\texttt{luca.marsilio@polimi.it} \\
	\And
	\href{https://orcid.org/0000-0003-3980-8296}
 {\includegraphics[scale=0.06]{orcid.pdf}\hspace{0.1mm}Davide Marzorati} \\
	Institute of Digital Technologies for\\
	Personalized Healthcare, Department \\
    of Technology and Innovation, Uni-\\
	versity of Applied Sciences and Arts of \\
    Southern Switzerland, Lugano, Switzerland.\\
    \And
	\href{https://orcid.org/0000-0003-2519-0720}
 {\includegraphics[scale=0.06]{orcid.pdf}\hspace{0.1mm}Matteo Rossi} \\
	Department of Electronics,\\
	Information, and Bioengineering\\
	Politecnico di Milano, Milan, Italy. \\
 \And
	\href{https://orcid.org/0000-0002-3365-580X}
 {\includegraphics[scale=0.06]{orcid.pdf}\hspace{0.1mm}Andrea Moglia} \\
	Department of Electronics,\\
	Information, and Bioengineering\\
	Politecnico di Milano, Milan, Italy. \\
 \And
	\href{https://orcid.org/0000-0002-6276-6314}
 {\includegraphics[scale=0.06]{orcid.pdf}\hspace{0.1mm}Luca Mainardi} \\
	Department of Electronics,\\
	Information, and Bioengineering\\
	Politecnico di Milano, Milan, Italy. \\
 \And
	\href{https://orcid.org/0000-0003-1791-6800}
 {\includegraphics[scale=0.06]{orcid.pdf}\hspace{0.1mm}Alfonso Manzotti} \\
	Hospital ASST FBF-Sacco,\\
	Milan, Italy. \\
 \And
	\href{https://orcid.org/0000-0003-3995-8673}{\includegraphics[scale=0.09]{orcid.pdf}\hspace{1mm}Pietro Cerveri} \\
	Department of Industrial and Information \\       Engineering, Università di Pavia, Pavia, Italy. \\
        Department of Electronics, Information, and \\ Bioengineering, Politecnico di Milano, Milan,  Italy.\\
}

\renewcommand{\shorttitle}{Cascade learning in multi-task encoder-decoder networks for concurrent bone segmentation and glenohumeral joint assessment in shoulder CT scans}

\hypersetup{
pdftitle={Cascade learning in multi-task encoder-decoder networks for concurrent bone segmentation and glenohumeral joint assessment in shoulder CT scans},
pdfsubject={cs.AI, cs.CV},
pdfauthor={Luca Marsilio, Davide Marzorati, Matteo Rossi, Andrea Moglia, Luca Mainardi, Alfonso Manzotti, Pietro Cerveri},
pdfkeywords={CT segmentation, Shoulder diagnostic, Deep learning, PSI-based intervention, Preoperative planning},
}

\begin{document}
\maketitle

\begin{abstract}
Osteoarthritis is a degenerative condition that affects bones and cartilage, often resulting in the formation of osteophytes, loss of bone density, and narrowing of joint spaces. Treatment options to restore normal joint function vary depending on the severity of the condition. This work introduces an innovative deep-learning framework processing shoulder CT scans. It features the semantic segmentation of the proximal humerus and scapula, the 3D reconstruction of bone surfaces, the identification of the glenohumeral (GH) joint region, and the staging of three common osteoarthritic-related conditions: osteophyte formation (OS), GH space reduction (JS), and humeroscapular alignment (HSA). The pipeline comprises two cascaded CNN architectures: 3D CEL-UNet for segmentation and 3D Arthro-Net for threefold classification. A retrospective dataset of 571 CT scans featuring patients with various degrees of GH osteoarthritic-related pathologies was used to train, validate, and test the pipeline. 
Root mean squared error and Hausdorff distance median values for 3D reconstruction were 0.22 mm and 1.48 mm for the humerus and 0.24 mm and 1.48 mm for the scapula, outperforming state-of-the-art architectures and making it potentially suitable for a PSI-based shoulder arthroplasty preoperative plan context. The classification accuracy for OS, JS, and HSA consistently reached around 90\% across all three categories. The computational time for the entire inference pipeline was less than 15~s, showcasing the framework's efficiency and compatibility with orthopedic radiology practice. The outcomes represent a promising advancement toward the medical translation of artificial intelligence tools. This progress aims to streamline the preoperative planning pipeline delivering high-quality bone surfaces and supporting surgeons in selecting the most suitable surgical approach according to the unique patient joint conditions.
\end{abstract}


\section{Introduction}
\label{sec:introduction}

Osteoarthritis (OA) is a degenerative condition affecting bones and cartilage, usually ending up in changes of the bony surfaces such as the formation of osteophytes, bone density loss, and the narrowing of joint spaces \cite{Kellgren1957, Ogawa2006, Thomas2016, Khazzam2020}. In the shoulder, the glenohumeral (GH) joint is formed by the glenoid (or humeral socket) and the humeral head, physiologically concentric to the glenoid. When primary OA develops, cartilage deterioration leads to a sensible reduction of the GH space, ultimately causing direct contact between the humeral head and its socket. This condition can progress up to complete impingement inducing inflammation, pain, and the reduction of the joint's range of motion. As OA severity increases, osteophytes might arise in the antero-inferior portion of the humeral head and extend downward \cite{Ogawa2006}. The constant bone rubbing can also lead the glenoid to flatten, along with further osteophyte progression to its boundaries \cite{Lo2021}. On a long-term basis, the overall process disrupts the GH joint functionality \cite{Lo2021}. Different treatment options can be pursued based on the severity of the condition and the osteophyte location. They span from rehabilitation and drug therapy to joint resurfacing and total arthroplasty, offering a spectrum of interventions tailored to individual needs \cite{Junker2016, Ward2018, Leung2020}. Moreover, a pathological humeroscapular alignment (HSA) further contributes to joint instability and osteoarthritis development \cite{Sassoon2013, Sabesan2014}. This condition, also called humeral eccentricity, involves the humeral head shift to the glenoid surface in the superoinferior and anteroposterior planes. Typically, it requires the surgical repositioning of the humeral head within its socket, following rehabilitation \cite{Kohles2007}. If the eccentricity is associated with osteoarthritis, shoulder arthroplasty may be considered an effective therapeutic option \cite{Buck2008, Kim2016}. In this situation, the assessment of the HSA before surgery improves the efficacy of the clinical procedure, and the rehabilitation plan after the operation can be tailored accordingly \cite{PetscavageThomas2014, Kleim2022}. Therefore, GH osteophyte severity, intra-articular space staging, and HSA are among the ultimate predictive factors for shoulder joint treatment. The ensemble of these conditions is crucial to identifying the most suitable surgical implant. Anatomical and reverse are the two main prosthetic devices used in shoulder replacement surgeries to restore joint function. An anatomical shoulder implant is designed to mimic the natural shoulder anatomy, while the reverse is specialized for individuals with severe osteoarthritis combined with irreparable rotator cuff damage. Since a strong humeral head eccentricity is strongly correlated to rotator cuff damage \cite{saupe2006association, goutallier2011acromio}, an eccentric HSA with a high degree of osteophyte size and GH joint space narrowing suggest a reverse approach, while concentric HSA and a physiological GH distance an anatomical prosthesis \cite{goetti2021biomechanics}.
Furthermore, an accurate alignment in primary arthroplasty intervention is related to a reduced need for revision procedures, enhancing the outcomes of such techniques \cite{watters2011analysis}. Personalized Surgical Instruments (PSIs) recently revealed decreased surgical time and a more repeatable neutral postoperative alignment, reducing also intra-operative complications \cite{Seon2016, noble2012value, Mattei2016}. PSIs for shoulder arthroplasty intervention are patient-specific cutting jigs reproducing the patient’s proximal humerus and scapula edges. Their production and deployment require access to digital three-dimensional (3D) volumes of the humerus and scapula, obtained through medical image processing \cite{Cerveri2017, lombardi2008patient}. These surfaces are needed to properly manufacture the disposable jigs, choose the prosthesis size, estimate the optimal cutting planes, and reduce the risk of poor fitting and associated loosening of the implants \cite{mandolini2022comparison}. Diagnostics and pre-operative computed tomography (CT) or magnetic resonance imaging (MRI) of the upper thorax are mandatory to identify osseous morphology, stage articular surface degeneration, characterize pathological deformations, and assess bone misalignment. Nevertheless, the presence of GH osteoarthritic-related pathological conditions results in highly irregular bone profiles, making the delineation of surface boundaries challenging even for expert radiologists \cite{Goud2008}. In addition, automated image processing and analysis still represent a critical issue in the orthopedic pipeline \cite{Khazzam2020, Maffulli2020, Ladd2021}. Deep learning-based tools, such as convolutional neural networks (CNNs), were explored in this field \cite{Isensee2021, Wang2021, Yeoh2021}. Encoder-decoder (E-D) architectures, such as the UNet and nnUNet, were effective in the identification of osseous regions and soft tissues in two-dimensional (2D) X-ray and volumetric (3D) scans \cite{Norman2018, Ambellan2019, Marzorati2020, Wang2021, Rossi2022, Florkow2022, Schnider2022, Isensee2021}. CNNs were also studied to stage osteoarthritis \cite{Thomas2020, Taghizadeh2021, Wang2022} and optimal treatment prediction, offering a valuable tool to assist clinicians in preoperative planning.\cite{Leung2020, Potty2023}. In this work, we propose an innovative deep learning framework designed for processing shoulder CT scans, aiming to streamline a PSI-based shoulder preoperative planning pipeline by delivering high-quality bone surfaces and support surgeons in selecting the most suitable surgical approach according to the unique patient joint conditions. It includes the semantic segmentation of the proximal humerus and scapula with the CEL-UNet model \cite{Rossi2022}, the 3D reconstruction of bone surfaces, the identification of the GH joint region, and the staging of three common osteoarthritic-related conditions: osteophyte formation (OS), glenohumeral joint space reduction (JS), and humeroscapular alignment (HSA).

\subsection{Literature background}

\subsubsection{Automated CT segmentation methods for PSI-based preoperative plan}

The success of PSI-based surgery relies on achieving sub-millimetric alignment between the patient’s bone surface and the jig footprint \cite{ogura2019high, anderl2016patient}. Consequently, the accuracy of bone segmentation methods is critical to ensuring precise instrumentation alignment with the patient’s anatomy. Traditional automated segmentation techniques, such as statistical shape models (SSMs) and region-growing algorithms, have been investigated for this purpose. A hierarchical acetabolum and femoral head SSM was proposed for the automatic segmentation of diseased hip \cite{yokota2013automated}. In \cite{lamecker20043d} and \cite{seim2008automatic}, 3D SSMs were deployed for pelvic bones segmentation from CT data, while in \cite{poltaretskyi2017prediction} the pre-morbid proximal humeral anatomy was predicted in patients with severe degenerative osteoarthritis or a fracture of the proximal humerus. Region-growing algorithms have also been utilized for surgical planning applications \cite{mendoza2012fast, sivewright1994interactive, sekiguchi1994interactive}. However, both SSMs and region-growing algorithms face significant challenges when dealing with complex anatomical structures and the variability often present in damaged joints. CNNs, such as the UNet \cite{ronneberger2015u} and the nn-UNet \cite{Isensee2021}, demonstrated the ability to overcome these limitations by effectively capturing intricate patterns and accommodating anatomical variability. They were successful in knee bone CT scans \cite{Marzorati2020, Zhang2022, Rossi2022}, vertebral bodies in CT images \cite{Klein2019, Suri2021, Schmidt2022}, and automatically produced tibia and femur volumes compliant for a PSI-based total knee arthroplasty intervention \cite{Marsilio2023}. 3D UNet architectures also systematically outperformed their 2D counterparts in processing volumetric scans. \cite{Klein2019, Marzorati2020, Schnider2022}.

\begin{figure}[!b]
    \centering
    \includegraphics[width=1\linewidth]{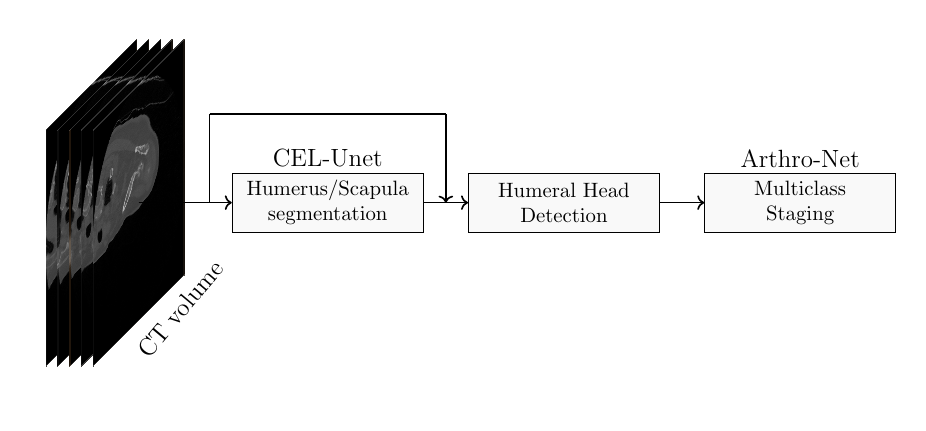}
    \caption{The proposed processing pipeline involving the volume segmentation to extract the humerus and the scapula (CEL-UNet model), the detection of the humeral head region in the original CT volume exploiting the two reconstructed surfaces, and finally the multiclass staging of the GH joint (Arthro-Net model), in terms of osteophyte staging (3 classes of increasing osteophyte size), GH joint space (3 classes of increasing severity) and humeral scapular alignment (2 classes).}
    \label{fig:cascade_pipeline}
\end{figure}

\subsubsection{Deep learning approaches to image-based joint clinical evaluation}
CNNs were proposed to stage cartilage osteonecrosis in the knee from X-ray images according to the Kellgren-Lawrence scoring, reporting results similar to intra-expert repeatability \cite{Thomas2020, Yeoh2021}.  In \cite{Uysal2021}, different pre-trained models, including ResNet and DenseNet, were compared to assess fracture/non-fracture conditions in X-ray images. A multi-task deep learning model was proposed for grading radiographic (weight-bearing anterior-posterior pelvic) hip osteoarthritis in 4368 patients \cite{Schacky2020}, disregarding other clinical conditions. In the shoulder, rotator cuff muscle degeneration was assessed using a CNN that analyzed CT scans from 95 patients, achieving accuracy comparable to that of expert raters \cite{Taghizadeh2021}. A regression CNN was used to estimate the critical shoulder angle \cite{Minelli2022} from anteroposterior shoulder X-ray images, with outcomes compatible with the clinical setting. A similar pipeline was also proposed to assess the most common causes of shoulder pain, such as proximal humeral fractures, joint dislocation, periarticular calcification, osteoarthritis, osteosynthesis, and joint endoprosthesis, on 2700 X-ray images \cite{Grauhan2022}. However, enhancing the shoulder evaluation for a patient-specific surgical plan requires a more comprehensive joint analysis, including osteophyte staging and humeroscapular alignment, often underestimated in the current literature. 

\subsection{Contributions}
Considering the preceding context, this study proposes a novel cascaded deep learning pipeline (Fig. \ref{fig:cascade_pipeline}), combining the semantic segmentation of proximal humerus and scapula, the 3D reconstruction of the two bones, and a multi-task multi-class classifier, predicting osteophyte staging, GH osteoarthritis severity, and HSA, concurrently. The paper's contributions can be summarized as:

\begin{itemize}
    \item the CEL-UNet segmentation quality assessment against state-of-the-art nnUNet architectures over a 571 CT scan dataset featuring several degrees of bone morphological deformations;
    \item the automatic extraction of the GH joint region in the CT volume to speed up the identification of the region where relevant joint pathological conditions are detected;
    \item the design of a novel CNN architecture (Arthro-Net) tailored for staging three GH osteoarthritic-related pathological conditions, concurrently;
    \item the deployment of an automated deep learning-based pipeline coupling two cascaded CNN architectures to simultaneously perform CT segmentation, 3D bone reconstruction, and GH pathology classification.  
\end{itemize} 

\section{Materials and Methods}
\label{sec:methodology}

\begin{figure}[!b]
   \centering
    \includegraphics[width=0.8\linewidth]{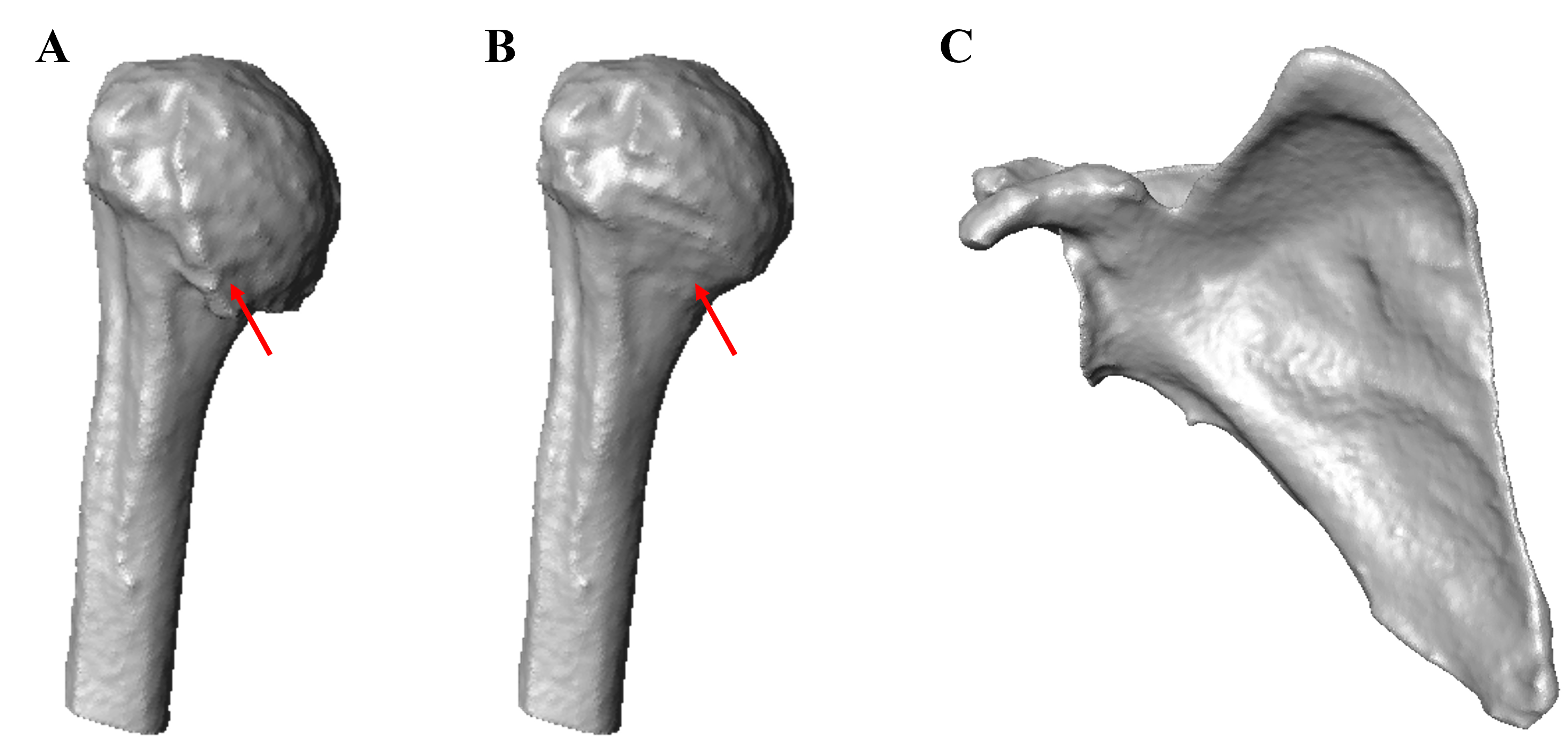}
    \caption{Example of bone volumes provided in the dataset A) original humerus morphology with osteophytes pointed out by the red arrow B) osteophyte-cleared humerus C) scapula.}
    \label{fig:stl_volumes}
\end{figure}

\subsection{Dataset description}
A dataset of 607 axial CT scans, acquired in DICOM standard format in the context of PSI-based preoperative planning for total or reverse shoulder arthroplasty intervention, was provided by MEDACTA International SA (Castel San Pietro, Switzerland). Patients, aged 74$\pm11$ years, were diagnosed with various clinical conditions like osteoarthritis, inflammatory arthritis, osteonecrosis involving the glenoid, and post-traumatic degenerative GH joint disease. Preliminary image analysis confirmed varying degrees of osteophytes, cartilage defects, misalignment of the humeral head, and GH impingement. All patients underwent shoulder arthroplasty between 2021 and 2022 (about 20\% anatomical implants and the remaining inverse implants). CT scans were acquired with different imaging equipment, mostly at 512$\times$512 pixels and 330 slices on average, with variable pixel size, ranging from 0.30 to 0.98~mm, and axial slicing, ranging from 0.30 to 2.5~mm. 
In addition to CT images, the dataset included 3D surface models of the proximal humerus and scapula in STL format. For each patient, two distinct humeral surfaces were provided: the original proximal humerus morphology and a modified version with osteophytes manually removed to represent a physiological humeral head anatomy (Fig.\ref{fig:stl_volumes}). These two surfaces were created to enhance the PSI-based preoperative planning process. The original morphology is essential for designing custom-cut implants and planning the contact areas between the cutting jigs. The osteophyte-cleared model is crucial for determining the appropriate prosthesis size and optimal positioning. Bony surfaces were generated by expert radiological operators from semi-automated CT segmentation with Mimics software (v.16.0) by Materialise (Leuven, Belgium). To enhance the reliability of these annotations, each surface was revised by one additional expert. STL volumes were then processed with 3D Slicer (v.5.2.2) to extract the segmentation labels to train, validate, and test the performance of the models.

\begin{figure}[!b]
    \centering
    \includegraphics[width=0.9\linewidth]{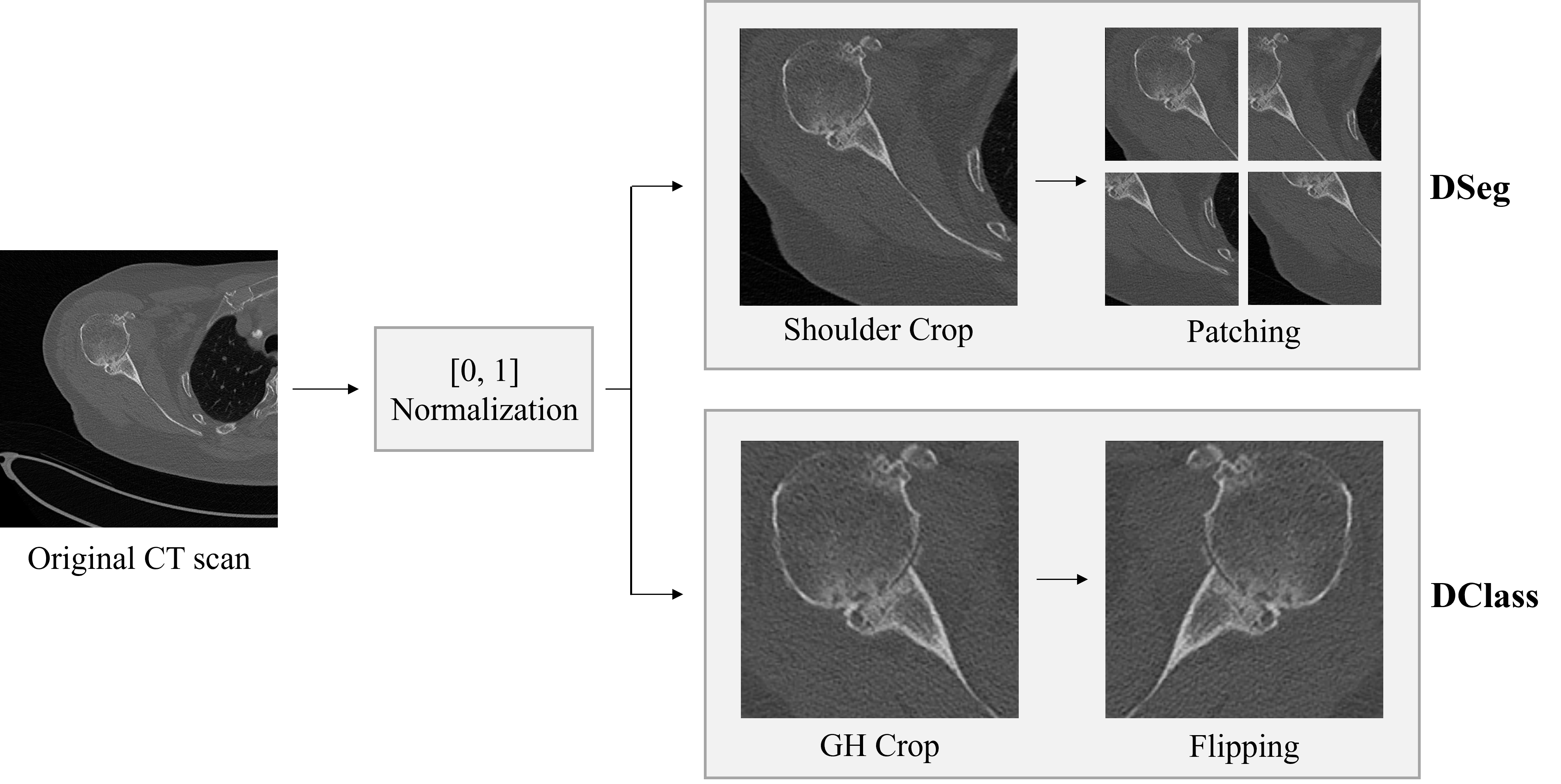}
    \caption{Segmentation (DSeg) and classification (DClass) training set generation pipeline. After CT normalization, DSeg is produced by cropping the volume to focus on the shoulder joint and patching the extracted sub-volumes, while for DClass a GH-centered crop was performed, and data augmentation was obtained by flipping the sub-volumes on the sagittal plane.}
    \label{fig:training_sets_generation}
\end{figure}

\subsection{Training set preparation}
Out of the initial 607 cases, 36 were excluded due to the presence of inner metal parts such as screws, implants, and plaques, which fell outside the scope of this study. The remaining 571 cases (297 males and 274 females, comprising 271 left shoulders and 300 right shoulders) were considered for the present investigation. Since CT scans originated from different scanning equipment, voxel intensities were normalized to improve the network's learning performance. Values, acquired in the Hounsfield Unit (HU) scale, were first clipped between -1024 and 2500 units, representing air to dense cortical bone \cite{Schneider1996}, then shifted to positive ranges, and normalized between 0 and 1 for consistency.
From this dataset, 85\% of the cases, totaling 481, were randomly selected for training and validation sets, including 410 and 71 volumes, respectively. The remaining 90 images defined the test set to evaluate the network performances. To address the two tasks, the dataset was duplicated into two sets: one for segmentation (DSeg), and one for classification (DClass) (Fig. \ref{fig:training_sets_generation}).
CT scans in the DSeg set were cropped in the axial, coronal, and sagittal view, to remove all the slices where the proximal humerus and scapula were unavailable. A patch-based method was then applied to increase training and validation instances, while preserving the original voxel resolution. Each cropped CT was patched into a different number of 160$\times$160$\times$160 sub-volumes, depending on their initial dimension and degree of overlapping (around 25\% of the size).
For the DClass set, CT scans were cropped to a single 160$\times$160$\times$160 sub-volume, ensuring that both the humeral head and the glenoid surface were included and centered on the GH joint area. This focused extraction was essential to minimize classification biases from surrounding tissues and organs and to reduce computational overhead. The GH-centered bounding box was extracted through an automated algorithm that builds upon the technique described in \cite{Cerveri2017}, originally focused on the proximal femoral head, that employs morphological analysis of the proximal humerus to delineate the desired region. The number of CT scans in the DClass set was finally doubled by flipping the originated sub-volumes in the sagittal plane. This technique effectively mirrored each joint, generating complementary and physiological counterparts, increasing the dataset number to enhance model generalization. 

\subsection{GH pathological conditions staging and data labeling}
\begin{figure}[!b]
    \centering
    \includegraphics[width=1\linewidth]{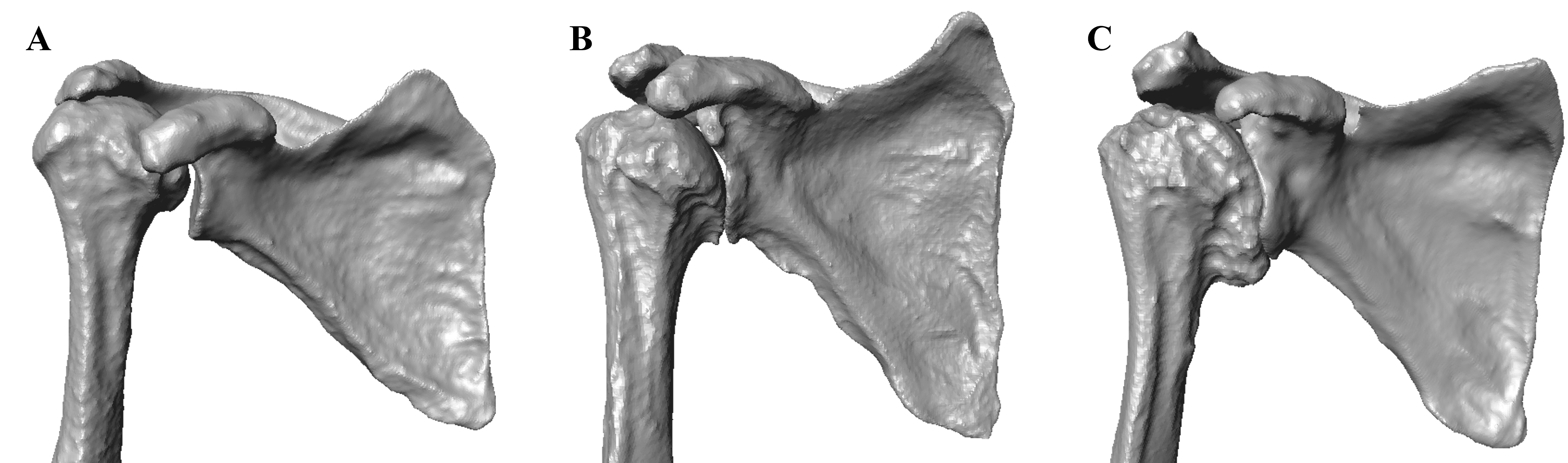}
    \caption{Different pathological conditions. A) Extreme eccentricity of the humeral head, B) Mild GH osteoarthritis, C) Severe GH osteoarthritis with large osteophyte presence.}
    \label{fig:gh_pathology_severity}
\end{figure}

The severity of humeral head osteophytes (OS) was staged into three classes based on the Samilson-Prieto grading system \cite{Elsharkawi2013, Habermeyer2017}, featuring increasing osteophyte size. Grade 0 denoted small-size ($s_o<$ 3~mm), grade 1 medium-size ($3<s_o<7$~mm), and grade 2 large-size ($s_o>$ 7~mm) osteophytes. Preliminary labeling of OS was automatically performed by computing the maximum distance (in millimeters) between the morphologic and osteophyte-cleared humerus reference surfaces for each case in the dataset and assigning the corresponding classification label. A radiological expert with more than 20 years of orthopedic clinical practice (A.M.) revised the final labeling of the OS. The GH joint space was tagged manually according to the Kellgren-Lawrence grading system \cite{Kellgren1957} under the supervision of the same radiological expert. Three classes were identified for each case depending on the residual GH joint space (JS), featuring physiological joint space (grade 0), slightly narrowed joint space (grade 1), and non-detectable joint space (grade 2). HSA was assessed by analyzing the displacement of the humeral head relative to the glenoid surface defining either eccentric-pathological or concentric-physiological humeral head alignments \cite{Kleim2022}. The frequency across classes within each diagnostic task is reported in Table \ref{tab:classification_labels}. Fig. \ref{fig:gh_pathology_severity} presents three shoulder joints from the dataset with different pathological conditions of the GH joint. In case A, acromiohumeral distance (i.e., the space between the humeral head and the acromion) was absent, indicating significant eccentricity in the HSA. However, joint space was still visible and only small-size osteophytes were present in the anteroinferior portion of the humeral head. In both cases B and C, the intra-articular spaces were missing, with the humeral head and glenoid making direct contact. This bone-on-bone rubbing led to the development of medium and large-size osteophytes, respectively. The humeral head in joint B was still concentric to its socket, while joint C showed an eccentric condition.

\begin{table}[!t]
\caption{GH osteoarthritic-related condition labeling. The first column describes the classification tasks, namely osteophyte size (OS), GH joint space (JS), and humeroscapular alignment (HSA). The second and third columns show the labeling criteria for each multi-class task and the provided index. The last column reports the frequency of each class in the corresponding task.}
\vspace{0.15cm}
\centering
{%
\begin{tabular}{cccc}
\toprule
        Classification Task & Criteria & Index & Frequency [\%] \\ 
 \midrule
        \multirow{3}{*}{\textbf{Osteophyte staging (OS)}}
        & $<$3~mm & 0 & 31.1 \\ 
        & 3-7~mm & 1 & 36.1 \\
        & $>$7~mm & 2 & 32.8 \\
 \midrule
        \multirow{3}{*}{\textbf{GH joint space (JS)}} & Physiological & 0 & 38.2 \\ 
        & Narrowed & 1 & 27.2 \\
        & Non-detectable & 2 & 34.6 \\
 \midrule
        \multirow{2}{*}{\textbf{HSA}} & Concentric & 0 & 56.1 \\ 
        & Eccentric & 1 & 43.9 \\
 \bottomrule
\end{tabular}}
\label{tab:classification_labels}
\end{table}

\subsection{Segmentation network: CEL-UNet}
\begin{figure}[!b]
    \centering
    \includegraphics[width=0.95\linewidth]{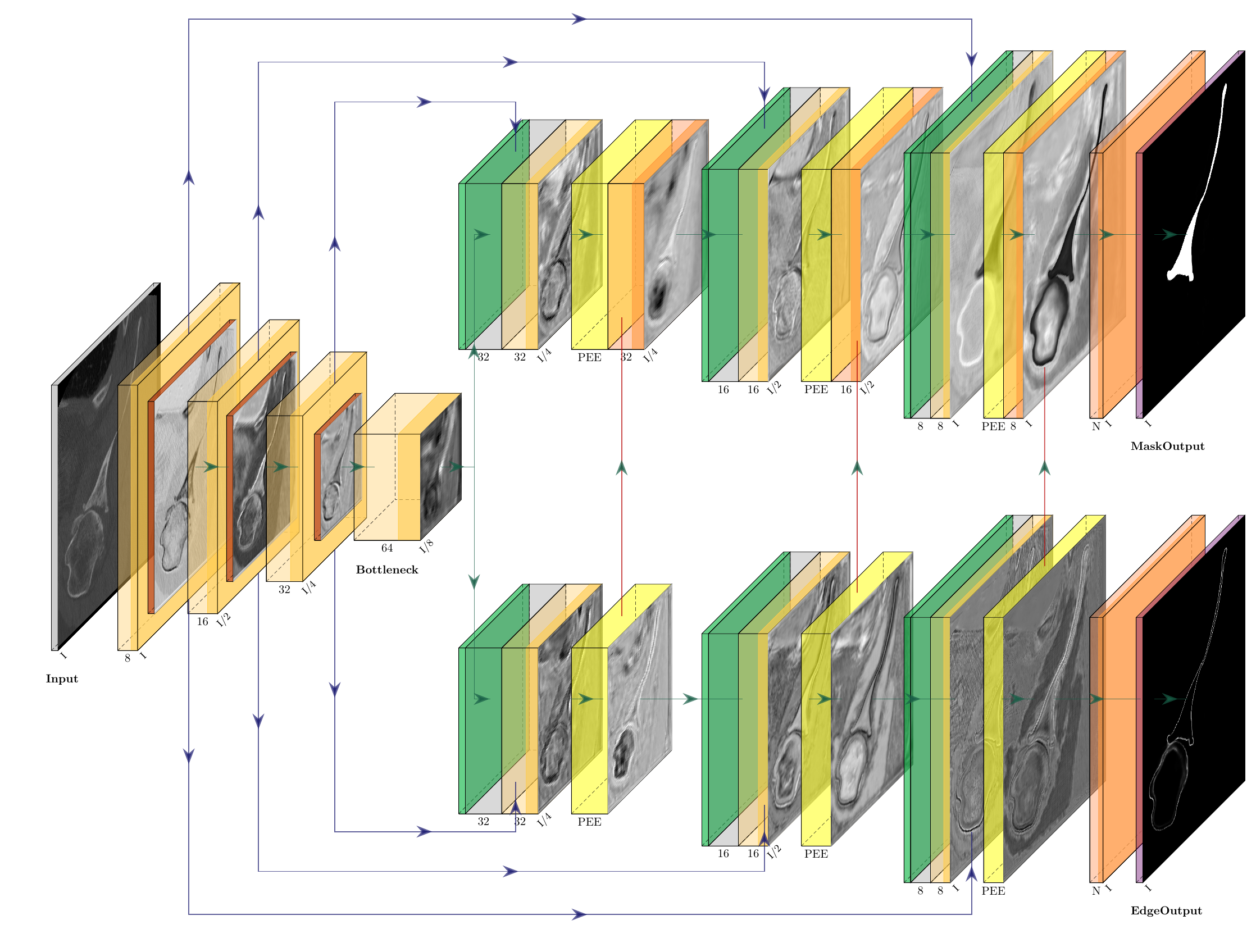}
    \caption{CEL-UNet architecture. The encoding path, on the left, extracts segmentation features at different resolutions, ending with a bottleneck. The decoder, on the right, is split into two parallel branches, devoted to region segmentation (RA) and edge detection (CA). The output of each CA decoding block is combined with the corresponding RA layer through vertical skip connections (red lines) to enhance semantic segmentation. Pyramidal edge extraction (PEE) modules (yellow boxes) are added after decoding blocks. The segmentation mask output is then used to reconstruct the bone surfaces using traditional matching cubes algorithm.}
    \label{fig:celunet}
\end{figure}

The CEL-UNet model was prior \cite{Rossi2022, Marsilio2023} designed and developed by our group to boost the segmentation quality on knee CT scans and enable preoperative planning in PSI-based total knee arthroplasty. The identification of severe pathological anatomy deformations and narrow joint space was enhanced by combining traditional semantic mask segmentation with boundary identification (Fig. \ref{fig:celunet}). In detail, the decoder was split into two parallel branches, one devoted to region segmentation (RA, region-aware branch) and the other addressing edge detection (CA, contour-aware branch). Vertical unidirectional skip connections were added between each CA branch processing block and its corresponding one in the RA branch, to integrate region and boundary-related features at increasing resolution. A specialized processing module was also implemented after each CA branch layer, using the pyramidal edge extraction (PEE) approach inspired by \citep{Wang2022}.
\noindent A custom loss function dynamically weighting RA and CA branch losses throughout the training process was deployed to better combine each contribution. It exploited the importance of the region boundary, by incorporating the distance-weighted map (DWM) factor into the formula. DWM was computed as the negative exponential function of the Euclidean distance transform (EDT) as: 

\begin{equation}
DWM_c = \left(1 + \gamma \cdot exp \left(-\frac{EDT_{c}}{\sigma}\right)\right)
\label{eq:DWM}
\end{equation}

\noindent where $c$ was the current label, while $\gamma$ and $\sigma$ were heuristic parameters. EDT assigned to each voxel the value of its distance from the closest voxel belonging to the boundary of the corresponding mask label. Hence, the distance-weighted cross-entropy loss $\mathcal{C}{_w}$ embedded the $DWM$ factor to raise the significance of contour and near-surface voxels as:

\begin{equation}
	{\mathcal C{_w}} = 
	-\sum_{c}^{C}\left(\sum^{N}\left(DWM_c \cdot y_{c} \cdot log({\hat{y}_{c}})\right)\right)
	\label{eq:Dce}
\end{equation}

\noindent where $y_{c}$ and $\hat{y}_{c}$, respectively the true and segmented voxels for the label $c$, were computed over the $N$ voxels. Thus, the loss of the region-aware branch was:

\begin{equation}
{\mathcal L_{RA}} = 1-(\alpha \cdot {\mathcal D} + (1-\alpha) \cdot {\mathcal{C}_w}) \\
\label{eq:SEG_loss}
\end{equation}

\noindent where $\alpha$ was a tunable parameter to modulate differently the two contributions during the training process and \textit{D} represents the dice formula.
In this work, CEL-UNet segmentation results were compared to the results obtained with the nnUNet proposed by \cite{Isensee2021} alternatively trained with two different loss functions, namely distance cross-entropy and focal.

\begin{figure}[!b]
    \centering
    \includegraphics[width=0.9\linewidth]{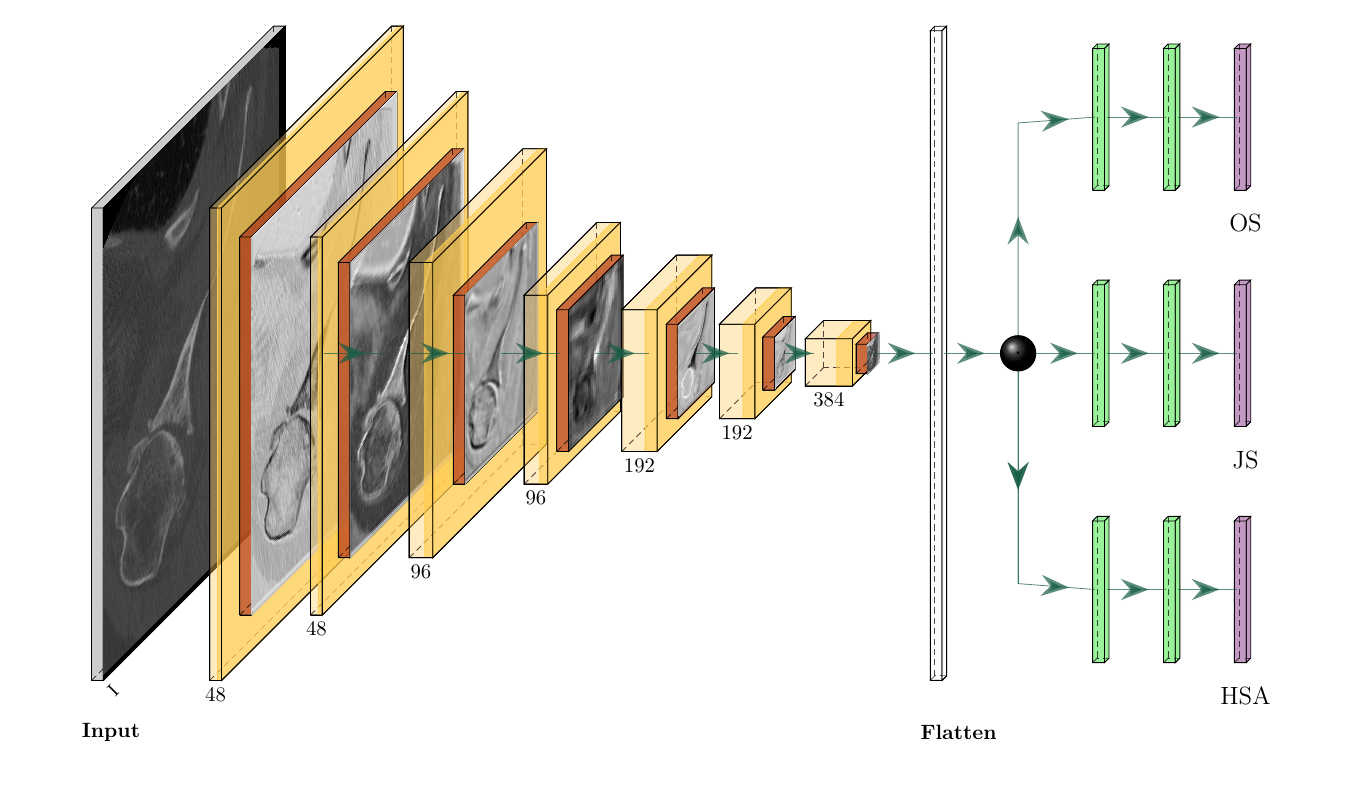}
    \caption{Arthro-Net architecture. Each processing block (yellow box) is featured by two sequences of convolutional with linear activation, batch normalization, and ReLU activation layers, mimicking the CEL-UNet encoder. Downsampling is performed through max pooling (orange box). The block ends with a flatten layer followed by three separate fully connected branches to specialize the network for the three classification tasks, namely osteophyte staging (OS), GH joint space (JS), and humeroscapular alignment (HSA).}
    \label{fig:arthro-net}
\end{figure}

\subsection{Classification network: Arthro-Net}
The Arthro-Net (Fig.\ref{fig:arthro-net}) is a novel CNN architecture designed for multi-task, multi-class staging of GH osteoarthritic-related conditions. The network processes GH-centered patches and classifies the severity of each pathology according to predefined categories: three classes for osteophyte size (OS), three classes for joint space narrowing (JS), and two classes for the humeroscapular alignment (HSA). The Arthro-Net encoder is a sequence of convolutional/downsampling blocks to extract features at decreasing resolution up to the bottleneck. Each processing block embeds two sequences of convolutional with linear activation, batch normalization, and ReLU activation layers. The filter size and stride length are 3$\times$3$\times$3 and 1$\times$1$\times$1, respectively. This configuration is consistent with the CEL-UNet feature extractor, which captures relevant osseous characteristics from CT images \cite{Rossi2022}. However, the number of feature maps in the Arthro-Net doubles every two processing blocks (Fig. \ref{fig:arthro-net}), to avoid unnecessary over-parametrization. Downsampling is performed through max-pooling layers with filter size and stride length of 2$\times$2$\times$2. According to state-of-the-art classification architecture \cite{Traore2018}, a flatten layer was placed after the bottleneck to reshape the data into a one-dimensional tensor while preserving the total number of elements. Three separate fully connected branches, characterized by two consecutive dense layers with input sizes of 256 and 32, and a ReLU activation function, were added to specialize the network for each classification task. Each branch was preceded by a dropout layer to prevent overfitting and reduce sparsity, due to the large number of unique connections at this stage. The dropout rate was set to 0.6. Lastly, the three outputs consisted of one softmax layer each, with three, three, and two neurons, respectively. Categorical cross-entropy was chosen as the loss function. To balance the uneven training label frequency for each task, shown in Table \ref{tab:classification_labels}, each loss function was weighted with a parameter $K_{c}$ to balance the representation of every class, following Eq.\ref{eq:inverse_frequency}:

\begin{equation}
	{\mathcal K{_c}} = \frac{\frac{1}{N{_c}}}{\sum_{i=1}^{C}(\frac{1}{N{_i}})} 
	\label{eq:inverse_frequency}
\end{equation}

\noindent where $c$, $N_{c}$, and $C$  were the current class for the specific classification task, the number of total cases for each class, and the number of classes, respectively. To further validate the novel architecture performances, an ablation test was performed to evaluate the dependency of the classification scores on the number of convolutional layers and feature maps. Two networks of increasing depth were analyzed, by training the model with either seven or eight processing blocks. Moreover, the initial number of feature maps was alternatively set to 16, 32, and 48 (the source code of the overall module is available at https://github.com/LucaMarsilio/AI-Shoulder.git).

\subsection{Network training and prediction pipeline}

The CEL-UNet and Arthro-Net were trained separately for the two tasks using Tensorflow framework (v.2.14). The learning rate for the Adam optimizer was fixed to 1e-04. Nevertheless, segmentation and classification were sequentially performed in the prediction step, combining the two networks with a cascade approach (Fig.\ref{fig:cascade_pipeline}). 
The CEL-UNet region identification output was processed by two custom algorithms. The first one is a marching cube-based method that generates the 3D humerus and scapula surfaces from the corresponding segmentation maps \cite{Cerveri2017}. The second creates a bounding box around the GH portion of the CT given the shoulder joint side (left or right) and the bone coordinates obtained in the previous step. The extracted sub-volume serves as the input for the Arthro-Net to classify osteoarthritic-related conditions. Training and testing procedures were performed on a 32-core CPU and NVIDIA A100-PCIE GPU with 40 GB RAM. Due to memory constraints, the batch size in the training was set to eight. Early stopping was used to prevent overfitting, stopping the training after 50 epochs with no improvement in validation loss. The total number of training epochs was 183.

\subsection{Segmentation and classification quality assessment}

The humerus and scapula segmentation performance on the test set was evaluated by comparing the CEL-UNet results (Dice and Jaccard indices) with those obtained by the state-of-the-art nnUNet\cite{Isensee2021}. The nnUNet was trained with two different loss functions, namely distance cross entropy (DCE) and focal (FOC) losses. Precision and recall were further computed to detect over- and under-segmentation errors. To evaluate 3D surface reconstruction accuracy, root mean squared error (RMSE) and Hausdorff distance were measured, capturing both average and maximum distances between predicted and reference volumes.
For classification tasks, accuracy, precision, recall, F1-score, and confusion matrices were utilized to determine the optimal Arthro-Net architecture. Time evaluations were conducted across the automated pipeline steps, encompassing CT segmentation, 3D reconstruction, GH identification, and pathological condition staging.
The statistical analysis of segmentation and 3D reconstruction outcomes was conducted using the Friedman test, followed by Wilcoxon Signed-Rank tests with Bonferroni correction for post-hoc analysis. This choice was driven by the features of the results: three paired sets with non-Gaussian distributions. Asterisk symbol (*) in Table \ref{tab:Segm_Results_AcrossNetworks} defines a statistically significant difference between the outcome distributions against the two other networks. Fig. \ref{fig:Segm_Results_Across_Networks} reports the unique differences between the CEL-UNet and the other models, a single asterisk (*) refers to a p-value $<$ 0.05, while three (***) to a p-value $<$ 0.001.

\begin{table}[!t]
\caption{Dice, Jaccard, precision, and recall median and IQR scores for humerus (\emph{above}) and scapula (\emph{below}) comparing the performances of the three networks (CEL-UNet, and DCE- and FOC-nnUNet). The asterisk symbol (*) defines a statistically significant difference between the outcome distributions against the two other networks.}
\vspace{0.15cm}
\centering
\resizebox{\textwidth}{!}{%
\begin{tabular}{ccccc}
\toprule
 \textbf{} & \multicolumn{4}{c}{\textbf{Humerus}} \\
 \midrule
        & \textbf{Dice} & \textbf{Jaccard} & \textbf{Precision} & \textbf{Recall} \\
    \midrule
DCE-nnUNet &		0.98 (0.97-0.99)&	0.97 (0.94-0.98)	&0.98 (0.97-0.99)	&0.98 (0.97-0.99) \\
FOC-nnUNet &		0.98 (0.97-0.99)&	0.96 (0.94-0.98)	&0.99 (0.98-0.99)	&0.97 (0.96-0.98)	\\
\textbf{CEL-UNet} &		\textbf{0.99* (0.98-0.99)}&	\textbf{0.98* (0.97-0.99)}	&\textbf{0.99 (0.98-0.99)}	&\textbf{0.99 (0.98-0.99)}	\\
\bottomrule
\end{tabular}}
\resizebox{\textwidth}{!}{%
\begin{tabular}{cccccc}
 \textbf{} & \multicolumn{4}{c}{\textbf{Scapula}} \\
 \midrule
        & \textbf{Dice} & \textbf{Jaccard} & \textbf{Precision} & \textbf{Recall} \\
    \midrule
DCE-nnUNet &		0.97 (0.96-0.98)&	0.95 (0.92-0.96)	&0.97 (0.95-0.98)	&\textbf{0.97* (0.96-0.98)} \\
FOC-nnUNet &		0.97 (0.96-0.98)&	0.94 (0.92-0.96)	&0.97 (0.96-0.98)	&0.97 (0.95-0.98)	\\
\textbf{CEL-UNet} &		\textbf{0.98* (0.97-0.98)}&	\textbf{0.95* (0.94-0.96)}	&\textbf{0.99* (0.99-0.99)}	&0.96 (0.95-0.97)	\\
\bottomrule
\end{tabular}}
\label{tab:Segm_Results_AcrossNetworks}
\end{table}

\section{Results}
\label{sec:results}

\subsection{Segmentation result comparison}

Focusing on the humerus (Table \ref{tab:Segm_Results_AcrossNetworks} and Fig. \ref{fig:Segm_Results_Across_Networks}),
Dice and Jaccard distributions achieved with the CEL-UNet were statistically higher than the other two networks. CEL-UNet also outscored DCE-nnUNet on precision and FOC-nnUNet on recall, respectively. Considering the scapula, statistically significant differences demonstrated the superiority of the CEL-UNet over state-of-the-art architectures in terms of Dice, Jaccard, and precision indices. The highest recall value was found with the DCE-nnUNet instead. A qualitative comparison of humerus and scapula segmentation across the three networks (Fig. \ref{fig:qualitative_segmentation_comparison}) highlighted the CEL-UNet’s ability to capture osteophyte contours (A, B) and narrowed scapular anatomy (C) accurately. In this example, DCE- and FOC-nnUNet did not recognize the bone spur (A, B), and FOC-nnUNet exhibited under-segmentation issues (black arrow).

\subsection{Humerus and scapula 3D reconstruction}
\begin{figure}[!b]
\centering
    {\includegraphics[width=0.49\textwidth] {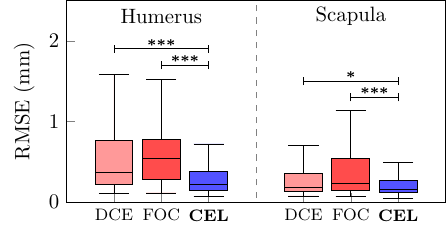} }
    \hskip 4pt
    {\includegraphics[width=0.49\textwidth]  {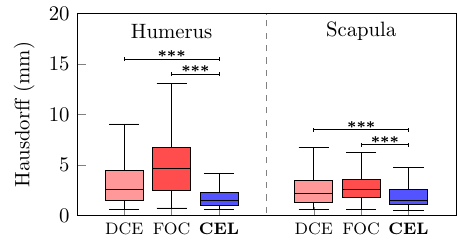} }
    \hskip 4pt

    \caption{Humerus and scapula 3D reconstruction errors. In detail, RMSE (\emph{left}) and Hausdorff distance (\emph{right}). Blue boxes depict the CEL-UNet results, light and dark red boxes DCE- and FOC-nnUNet scores. Asterisks (*, ***) report a statistically significant difference between CEL-UNet and other networks ($p<$0.05 and $p<$0.001, respectively).}
  \label{fig:3D_Results_Networks} 
\end{figure}

The CEL-UNet achieved global RMSEs of less than 1~mm for both 3D reconstructed surfaces (Fig. \ref{fig:3D_Results_Networks}). Median and IQR values were 0.22 (0.15-0.38)~mm, 0.37 (0.22-0.77)~mm, and 0.55 (0.29-0.78)~mm for the humerus, and 0.16 (0.12-0.27)~mm, 0.19 (0.13-0.36)~mm, and 0.24 (0.15-0.54)~mm for the scapula with CEL-UNet, DCE-nnUNet, and FOC-nnUNet, respectively. CEL-UNet statistically outscored the other networks.
Statistical evidence was confirmed with Hausdorff distance errors, reporting median and IQR of 1.48 (0.97-2.26)~mm, 2.56 (1.47-4.51)~mm, and 4.68 (2.50-6.73)~mm for the humerus, and 1.48 (1.13-2.56)~mm, 2.22 (1.31-3.48)~mm, and 2.63 (1.84-3.60)~mm for the scapula. The qualitative examination of the 3D humerus and scapula reconstructed surfaces for a test set case example (Fig. \ref{fig:3D_volume_comparison}) revealed the CEL-UNet proficiency in the reconstruction of irregular humeral head osteophytes (circled black area A), missed by DCE- and FOC-nnUNet.
All three networks faced difficulties in accurately reconstructing the scapula, particularly around the superior angle. This region's highly variable and often sub-millimeter thickness poses a challenge for segmentation at standard voxel resolutions. Despite this, the CEL-UNet delivered superior reconstruction quality, especially in the subscapular fossa.

\subsection{Arthro-Net: classification results of the three pathologic conditions}
\begin{figure} [!t]
    \centering
     \includegraphics[width=0.85\textwidth]{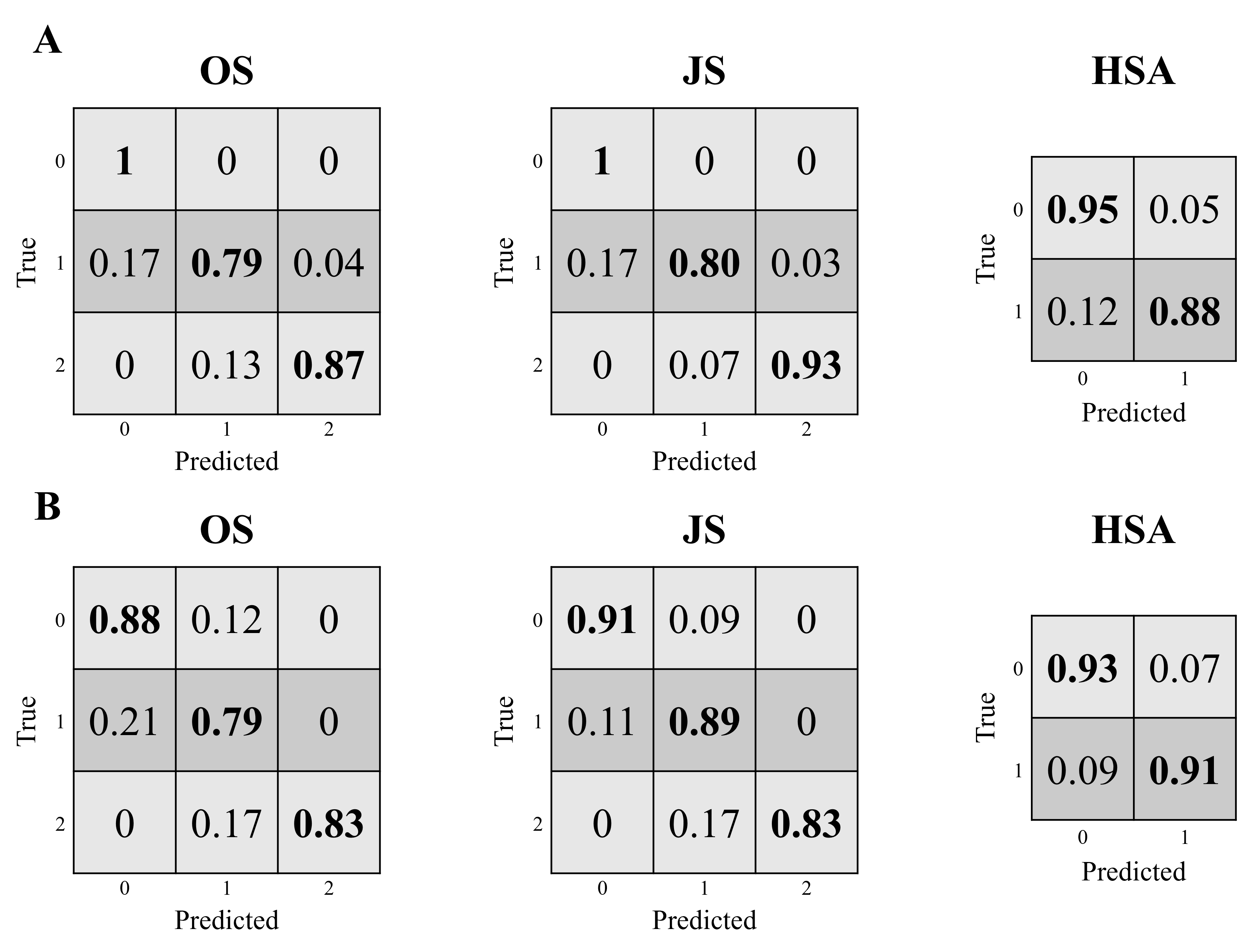}
   \caption{
   Confusion matrices of the two best Arthro-Net configurations are presented in Table \ref{tab:classification_results} for the three classification tasks (OS, JS, and HSA). A) Arthro-Net with seven processing blocks and 48 starting feature maps. B) Arthro-Net with eight processing blocks and 32 starting feature maps. 
   }
    \label{fig:CM_normalized}
\end{figure}

The ablation study on Arthro-Net's key hyperparameters (number of processing blocks and feature maps), revealed two optimal design configurations (Table \ref{tab:classification_results}). The first one, named $A$(7, 48), featuring seven processing blocks and 48 initial feature maps (crf. Fig. \ref{fig:arthro-net}), showed globally the best accuracy, precision, recall, and F1-score scores, ranging from 0.88 to 0.95 across the three classification tasks. Only the F1-score for HSA (0.88) was less than the corresponding score (0.90) in the design with 32 feature maps. The second one, $B$(8, 32), showed scores ranging from 0.83 to 0.94, greater than the other two competing configurations, while comparable with the previous design $A$.
The further analysis, performed on the normalized confusion matrices computed for $A$ and $B$ designs (Fig. \ref{fig:CM_normalized}), was resolutive in choosing the optimal Arthro-Net configuration. The osteophyte size staging was better achieved with design $A$, correctly predicting each small-size bone spur and outscoring design $B$ for the other sizes. Likewise, a perfect GH physiological joint space classification was obtained with design $A$, and non-detectable joint space was better identified as well. Design $B$ instead prevails in the identification of narrow joint space and the detection of eccentric humeroscapular alignment. Concentric condition staging was better achieved with design $A$. According to such results, the final Arthro-Net was configured according to design $A$. As expected the OS recall was more accurate when detecting small ($<$3~mm) and large ($>$7~mm) osteophytes than osteophytes of intermediate size (3-7~mm) (Fig. \ref{fig:CM_normalized}). Likewise, the recall results of both physiological (100\%) and non-detectable (93\%) JS conditions were more reliable than the recall value of narrowed (80\%) joint space. The classification of humeral head eccentricity was a little poorer (88\%) than concentricity (95\%). 

\begin{figure}[!t]
    \centering
    \includegraphics[width=1\textwidth]{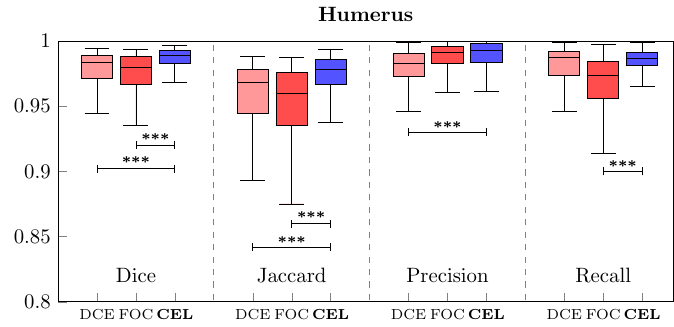}
    \par
    \vspace{0.25cm}
    \includegraphics[width=1\textwidth]{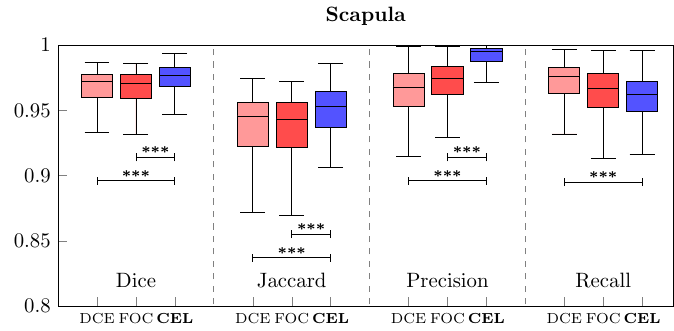}
    \caption{Boxplots of Dice, Jaccard, precision, and recall for humerus (\emph{above}) and scapula (\emph{below}). Blue boxes depict the CEL-UNet results, light and dark red boxes the nnUNet \cite{Isensee2021} trained with distance cross entropy (DCE) and focal (FOC) losses, respectively. Asterisks (***) report a statistically significant difference between the CEL-UNet and the other networks ($p<$ 0.001).}
    \label{fig:Segm_Results_Across_Networks}
\end{figure}

\begin{figure}[!t]
    \centering
     \includegraphics[width=1\textwidth]{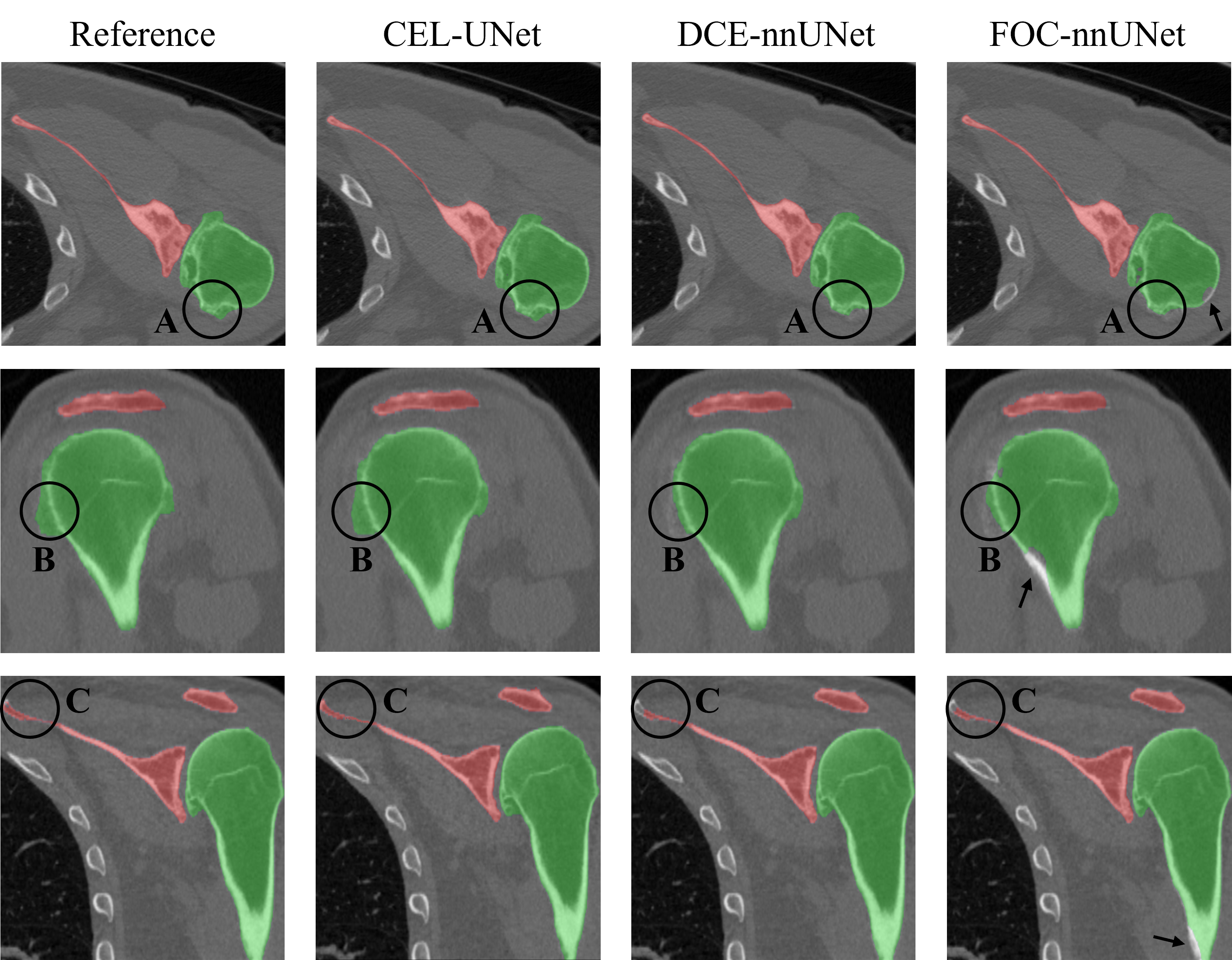}
   \caption{Qualitative comparison of the humerus (green) and scapula (red) segmentation for a test set case example against the reference label. Reference labels are in the first column, CEL-UNet outputs in the second, while DCE- and FOC-nnUNet segmentations are in the third and fourth, respectively. Circled areas display segmentation errors of the UNet models in identifying humeral head osteophytes (A and B) and the narrow scapular superior angle (C).}
    \label{fig:qualitative_segmentation_comparison}
\end{figure}

\subsection{Deployment of the overall pipeline and inference speed}
To validate the potential of translating the developed pipeline in clinical practice, the computational time of each operation was measured across the whole test set (Table \ref{tab:time_metrics}). This included CT segmentation using the CEL-UNet, 3D reconstruction, glenohumeral region identification, and diagnostic classification using the Arthro-Net. Among these, segmentation, encompassing CT loading, preprocessing, and postprocessing, emerged as the most time-consuming step, with a median time of 9.2~s. Proximal humerus reconstruction time outpaced that of scapula reconstruction, taking 1.0~s (range: 0.65-1.7) versus 3.6~s (range: 2.9-4.6), owing to the smaller size of the analyzed bone segment. The humeral head patching algorithm time to identify the glenohumeral region was neglectable, while the classification took about 1~s.
The cumulative median time for all processes amounted to 14.8~s, confirming a high-speed throughput underscoring the efficiency of the proposed cascaded multi-task approach making it potentially suitable for clinical diagnostic practice. 

\begin{figure} [!t]
    \centering
     \includegraphics[width=0.9\textwidth]{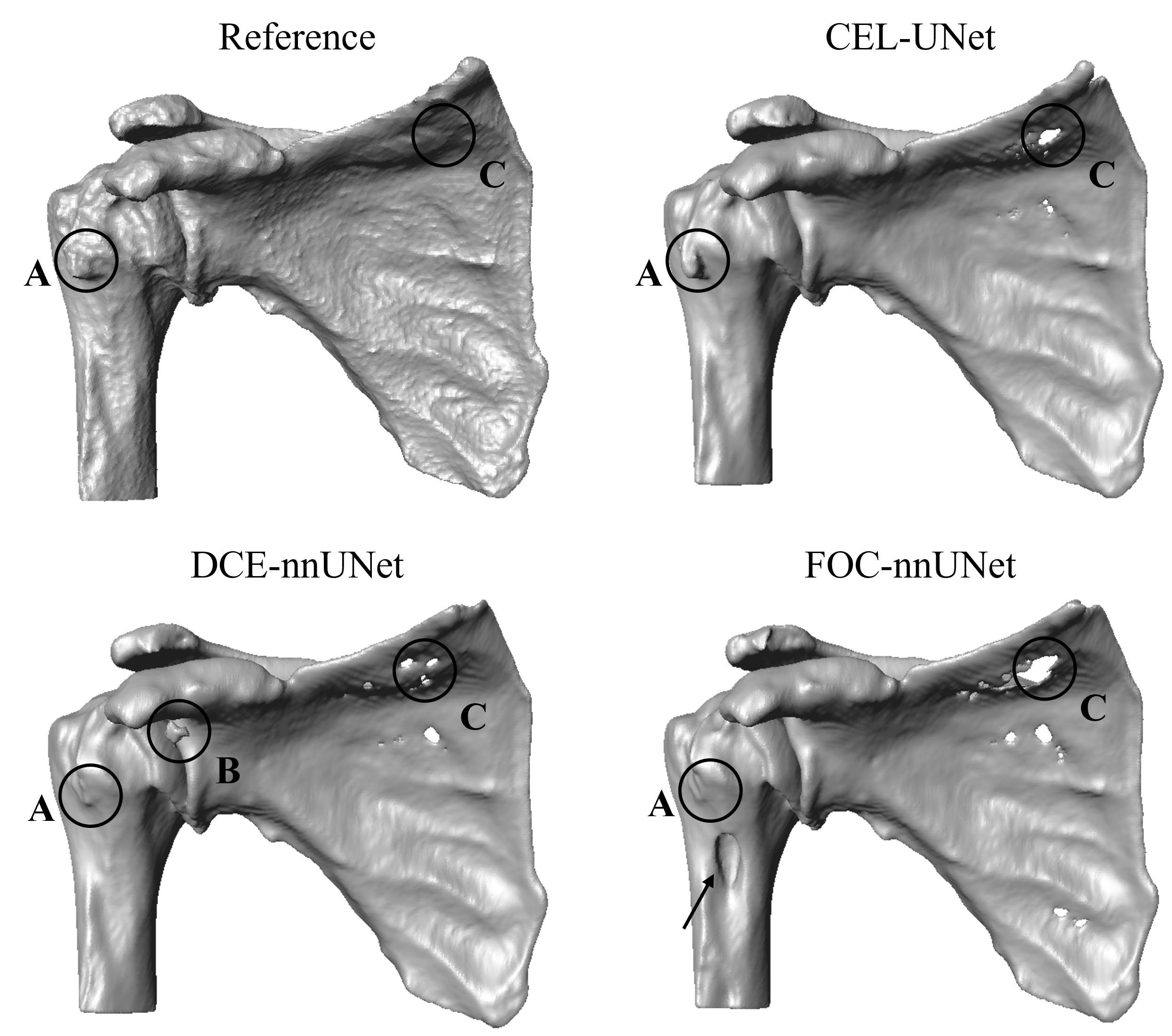}
   \caption{
   Qualitative 3D reconstruction comparison of the humerus and scapula for a test set case example against the reference label. A) irregular humeral head osteophyte reconstruction comparison B) scapula osseous portion wrongly identified by the DCE-nnUNet C) scapula superior angle reconstruction. The black arrow points out the error in the humerus shaft reconstruction from the FOC-nnUNet.}
    \label{fig:3D_volume_comparison}
\end{figure}

\begin{table}[!t]
\caption{Accuracy, precision, recall, and F1-score results for the three GH conditions classified, including osteophyte size (OS), joint space (JS), and humeroscapular alignment (HSA) with an increasing number of feature maps (FM) in the first processing block (16, 32, and 48), and comparing two Arthro-Net architectures featuring seven (\emph{above}) and eight (\emph{below}) total processing blocks.}
\vspace{0.15cm}
\centering
\resizebox{\textwidth}{!}{%
\begin{tabular}{ccccccccccccc}
\toprule
& \multicolumn{12}{c}{\textbf{7 Processing Blocks}} \\
\midrule
\textbf{FM} & \multicolumn{3}{c}{\textbf{Accuracy}} & \multicolumn{3}{c}{\textbf{Precision}} & \multicolumn{3}{c}{\textbf{Recall}} & \multicolumn{3}{c}{\textbf{F1-Score}} \\
\midrule
& OS & JS & HSA & OS & JS & HSA & OS & JS & HSA & OS & JS & HSA \\
\midrule
16 & 0.70 & 0.75 & 0.84 & 0.68 & 0.76 & 0.92 & 0.69 & 0.74 & 0.75 & 0.69 & 0.72 & 0.83 \\
32 & 0.84 & 0.89 & 0.90 & 0.84 & 0.90 & 0.91 & 0.83 & 0.89 & 0.90 & 0.83 & 0.89 & 0.91 \\
\textbf{48} & \textbf{0.89} & \textbf{0.91} & \textbf{0.91} & \textbf{0.89} & \textbf{0.92} & \textbf{0.95} & \textbf{0.88} & \textbf{0.91} & \textbf{0.88} & \textbf{0.89} & \textbf{0.91} & \textbf{0.91} \\
\midrule
\end{tabular}}
\resizebox{\textwidth}{!}{%
\begin{tabular}{ccccccccccccc}
& \multicolumn{12}{c}{\textbf{8 Processing Blocks}} \\
\midrule
\textbf{FM} & \multicolumn{3}{c}{\textbf{Accuracy}} & \multicolumn{3}{c}{\textbf{Precision}} & \multicolumn{3}{c}{\textbf{Recall}} & \multicolumn{3}{c}{\textbf{F1-Score}} \\
\midrule
& OS & JS & HSA & OS & JS & HSA & OS & JS & HSA & OS & JS & HSA \\
\midrule
16 & 0.78 & 0.73 & 0.87 & 0.79 & 0.71 & 0.97 & 0.78 & 0.72 & 0.77 & 0.78 & 0.69 & 0.86 \\
\textbf{32} & \textbf{0.84} & \textbf{0.88} & \textbf{0.92} & \textbf{0.85} & \textbf{0.89} & \textbf{0.94} & \textbf{0.83} & \textbf{0.88} & \textbf{0.92} & \textbf{0.84} & \textbf{0.88} & \textbf{0.93} \\
48 & 0.80 & 0.81 & 0.88 & 0.81 & 0.83 & 0.93 & 0.80 & 0.81 & 0.83 & 0.80 & 0.80 & 0.88 \\
\bottomrule
\end{tabular}}
\label{tab:classification_results}
\end{table}

\begin{table}[!t]
\caption{Time spent for each step in the pipeline, including CT segmentation, humerus and scapula 3D reconstruction, and GH-related condition staging. Tests were run on a 32-core CPU, and NVIDIA A100-PCIE GPU with 40 GB RAM. The same tasks were tested on a 12th Gen Intel(R) Core(TM) i7-12850HX, 2.10 GHz, 32.0 GB of RAM, and took on average 95 seconds.}
\vspace{0.15cm}
\centering
\resizebox{\textwidth}{!}{%
\begin{tabular}{c c c c c}
 \midrule
        & \textbf{Segmentation} & \textbf{3D Reconstruction} & \textbf{Classification} & \textbf{Overall} \\
    \midrule
\textbf{Time [s]} &	9.2 (7.7-11.6)	&	4.6 (3.6-6.4) & 0.9 (0.6-1.2) & 14.8 (11.7-17.1)\\
\bottomrule
\end{tabular}}
\label{tab:time_metrics}
\end{table}

\section{Discussion}
\label{sec:discussion}

In recent years, the use of deep learning algorithms for biomedical image processing has gained considerable attention in several key areas, including semantic segmentation for 3D anatomy reconstruction, morphological lesion analysis, and disease classification for clinical staging to aid diagnostics. However, much of the current literature is highly specialized, frequently focusing on just one of these domains at the time. In this work, a comprehensive automated framework for shoulder CT scan processing was developed and tested on a 571 retrospective dataset. It integrates the proximal humerus and scapula segmentation and 3D reconstruction, with the classification of three GH joint pathological conditions at various severity degrees. The outcomes showcased median root mean squared errors less than 0.5~mm for both bones, even in case of severe osteoarthritis, extensive morphological deformation, and inter-surface impingement (cfr. Fig. \ref{fig:Segm_Results_Across_Networks}). The CEL-UNet segmentation results outscored state-of-the-art nnUNet models (crf. Fig. \ref{fig:3D_Results_Networks}) \cite{Isensee2021}, revealing an advantage on severe pathological joints (Fig. \ref{fig:qualitative_segmentation_comparison}, \ref{fig:3D_volume_comparison}). A novel CNN (Arthro-Net) was further proposed to achieve an automated staging of the GH osteophyte size, joint space, and the relative position between the humeral head and the glenoid surface, following the CEL-UNet-based segmentation. 
The ablation study allowed us to identify the optimal balance between network depth and feature map size. Configurations with fewer feature maps and depth often lack the complexity needed to accurately capture detailed anatomical features, leading to poorer performance. Conversely, deeper networks with more feature maps may suffer from overfitting. The optimal results from the ablation study delivered a configuration that strikes a balance between model complexity and accuracy, mitigating both under and overfitting. The classification accuracy was around 90\% for all three class sets. 
A deterministic automated algorithm for the GH-centered patch extraction was implemented following shoulder segmentation, diverging from the conventional approach of integrating an additional deep learning model such as Mask-R-CNN, YOLO, or analogous object detection networks \cite{Anantharaman2018, Liu2023}. This approach streamlined the computational workflow and minimized potential errors introduced by inter-model dependencies.

\subsection{Literature overview and limitations}
Despite the recent advancements in automated image processing, several limitations persist in the existing literature, impacting the robustness, scalability, and clinical translation of the proposed methods. Many studies rely on relatively small datasets \cite{Schnider2022, Wong2023, kim2024development}, raising concerns about the generalizability of their approaches, not adequately representing the variability found in real-world clinical scenarios. Other works \cite{Liu2020, Noguchi2020} analyze CT images from patients who are not engaged in preoperative planning contexts, resulting in datasets with less severe joint damage. This can lead to reduced scalability when applied to cases with significant pathological conditions. Additionally, a reported mean 3D reconstruction error exceeding 1~mm, as in \cite{Wong2023, huang2022glenohumeral}, renders a method less suitable for PSI-based preoperative planning frameworks, where high reconstruction accuracy is crucial \cite{ogura2019high, anderl2016patient}. In \cite{huang2022glenohumeral}, the GH joint was reconstructed by deploying a SSM yielding to a mean surface RMSE of the humerus and scapula around 2~mm. A UNet-based glenoid segmentation model was trained and tested with a 237 CT scan dataset featuring patients with unilateral shoulder dislocations achieving an average intersection-over-union score of 0.96 \cite{zhao2023glenoid}. Nearly 100 MRIs were processed by AlexNet and UNet-based segmentation models to extract humerus, humeral head, and articular bone segmentation, reaching a positive predictive value of dice coefficient between 0.94 and 0.97 \cite{wang2021convolutional}.
The GH joint deep learning-based analysis has been reported to address different acquisition modalities, pathological conditions, and shoulder morphological parameters. 
In \cite{Sezer2020}, the authors reported a CNN to automatically extract discriminative features from proton density MRI sequences, achieving about 98\% success rate in classifying normal, edematous, and Hill-Sachs lesion humeral heads. However, the geometric relation between the humeral head and the glenoid surface was not investigated. A set of bone morphological parameters, including the glenoid inclination and the shoulder critical angle, were automatically assessed on about 1250 anteroposterior shoulder radiographs using a UNet model \cite{Shariatnia2022}. Nonetheless, the staging of potential osteoarthritic conditions of the GH was not computed. Finally, ensemble deep learning models were proposed to process X-ray images and assess fracture/non-fracture conditions in the shoulder area, featuring an accuracy of nearly 84\% \cite{Uysal2021}.

\subsection{Translational contributions and work limitations}
From a translation point of view, we highlight two main potential contributions of this work. Firstly, the achieved median root mean squared errors below 0.5~mm renders the segmentation tool potentially suitable to be integrated into a PSI-based planning pipeline, delivering radiologists high-quality bone surfaces within a computation time-frame that aligns with clinical requirements \cite{Marsilio2023}. Secondly, the three GH joint condition staging may support surgeons selecting the most suitable implant solution among the available ones. For instance, the proximal migration of the humeral head could suggest a large rotator cuff tear discouraging an anatomical prosthesis as well as advocating a reverse total shoulder replacement at the same time. The absence of osteophytes and a concentric humeral head could suggest more conservative solutions, including the GH joint resurfacing \cite{saupe2006association, goutallier2011acromio}. Likewise, based on the predicted osteophyte staging and the presence of a concentric humeral head, there arises a compelling argument in favor of GH joint resurfacing as a potentially superior alternative to complete anatomical shoulder replacement \cite{goetti2021biomechanics}.

It is essential nonetheless to interpret the findings of this study within the context of its limitations. Although traditional descriptions of humeral head eccentricity (subluxation) include posterior, anterior, inferior, and superior directions \cite{Kleim2022}, this study focused on a single binary HSA condition, restricting the comprehensive assessment of glenohumeral conditions. This choice was due to the retrospective nature of the study, involving a dataset with multiple cases of degenerative shoulder arthritis. Additionally, while the dataset included nearly 600 patients, it was biased towards an older demographic, with a median age of 74 years. This reflects a common trend in studies of degenerative shoulder conditions, where older patients, who are more susceptible to bone deterioration and decreased turnover \cite{von2016failure}, are often over-represented. To enhance the scalability and applicability of the segmentation module, future work should aim to include a more diverse patient population, but also cases with implants, plates, and screws to allow the deployment in preparation for revision procedures. The high throughput presented in Table \ref{tab:time_metrics} is achieved with high performance and dedicated hardware, not always available in clinics. Nevertheless, the same pipeline was tested on a laptop featuring 12th Gen Intel(R) Core(TM) i7-12850HX, 2.10 GHz, 32.0 GB of RAM, and equipped with an NVDIA RTX A200 8GB video card and completed in around 95~s. Finally, validating this framework with clinicians and surgeons will be essential to assess its practical advantages and effectiveness in real-world clinical scenarios.

\section{Conclusions}
\label{sec:conclusions}
We have introduced a comprehensive AI-based automated pipeline for reconstructing the two primary bones of the shoulder and predicting three distinct diagnostic conditions of the GH joint. While existing literature offers valuable insights, our study stands out for two key innovative contributions: firstly, our fully automated framework enables the concurrent proximal humerus and scapula 3D reconstruction and the staging of three major glenohumeral joint pathological conditions. This tool has the potential to be further validated within a PSI-based preoperative planning pipeline. It delivers high-quality bone surface models and a comprehensive assessment of the joint condition, that may assist surgeons in the selection of the most suitable surgical approach (arthroplasty or resurfacing) and implant type (anatomical or reverse), thereby streamlining the decision-making process. Secondly, our results derive from a larger test set compared to those in similar studies \cite{carl2024shoulder, Wong2023, Liu2020, Zhao2023}, enhancing the robustness and reliability of our findings.

\section*{CRediT authorship contribution statement}

\noindent \textbf{Luca Marsilio}: Conceptualization, Software, Data curation and analysis, Writing – original draft. \textbf{Davide Marzorati}: Formal analysis, Methodology. \textbf{Matteo Rossi}: Software, Visualization. \textbf{Andrea Moglia}: Investigation, Writing review \& editing. \textbf{Luca Mainardi}: Resources, Writing review \& editing. \textbf{Alfonso Manzotti} Data curation, Investigation. \textbf{Pietro Cerveri}: Conceptualization, Project administration, Funding acquisition, Supervision, Writing – review \& editing.

\section*{Declaration of interest statement}
\noindent The authors declare that they have no known competing financial interests or personal relationships that could have appeared to influence the work reported in this paper.

\section*{Ethical background}
\noindent Image data were retrospectively available in anonymized form by Medacta International SA (Castel San Pietro, Switzerland).

\section*{Acknowledgment}
\noindent This work was supported by the program PON-FESR 2014-2020 - European Regional Development Fund, ARIA-2020-403, ID 138.360.852, and by PNRR-PE, Future Artificial Intelligence Research (FAIR) - Italian Ministry of University and Research. The authors thank Medacta International SA (CH) for providing data.

\bibliographystyle{unsrtnat}
\bibliography{review} 

\end{document}